\documentclass[10pt,twocolumn]{article}
\usepackage{cite}
\usepackage{amsmath,amssymb,amsfonts,amsthm}
\usepackage{multicol}
\usepackage{caption}
\usepackage{graphicx}
\usepackage{csquotes}
\usepackage{todonotes}

\def\BibTeX{{\rm B\kern-.05em{\sc i\kern-.025em b}\kern-.08em
    T\kern-.1667em\lower.7ex\hbox{E}\kern-.125emX}}
\usepackage{hyperref}
\usepackage[margin=1.75cm]{geometry}
\usepackage[hpos=0.72\paperwidth,vpos=0.97\paperheight,angle=0,scale=0.8]{draftwatermark}
\SetWatermarkText{Preprint}
\SetWatermarkLightness{0.5}
\SetWatermarkText{Preprint of preliminary work}

\title{Optimally Reliable \& Cheap Payment Flows on the Lightning Network}
\author{Rene Pickhardt \& Stefan Richter\thanks{The authors have contributed to this work in equal measure. The order merely reflects the fact that this line of research was initiated by Rene Pickhardt.}}
\begin{document} 
\maketitle

\begin{abstract}
  Today, payment paths in Bitcoin's Lightning Network are found by searching for shortest paths on the fee graph.
  We enhance this approach in two
  dimensions. Firstly, we take into account the probability of a payment
  actually being possible due to the unknown balance distributions in the
  channels. Secondly, we use minimum cost flows as a proper generalization of
  shortest paths to multi-part payments (MPP).

  In particular we show that under plausible assumptions about the balance
  distributions we can find the most likely MPP for any given set of senders, recipients and amounts by solving for a (generalized) integer
  minimum cost flow with a separable and convex cost function. Polynomial time
  exact algorithms as well as approximations are known for this optimization
  problem.
  %We evaluate our approach on a reference implementation of an exact algorithm as well as on a surprisingly efficient and precise heuristic that we found by accident.

  We present a round-based algorithm of min-cost flow computations for
  delivering large payment amounts over the Lightning Network. This algorithm
  works by updating the probability distributions with the information gained
  from both successful and unsuccessful paths on prior rounds. In all our
  experiments a single digit number of rounds sufficed to deliver payments 
  of sizes that were close to the total local balance of the sender.
  Early experiments indicate that our approach increases the size of payments that can be reliably delivered by several orders of magnitude compared to the current state of the art.

  We observe that finding the cheapest multi-part payments is an NP-hard problem
  considering the current fee structure and propose dropping the base fee to make
  it a linear min-cost flow problem. Finally, we discuss possibilities for
  maximizing the probability while at the same time minimizing the fees of a
  flow. While this turns out to be a hard problem in general as well --- even in
  the single path case --- it appears to be surprisingly tractable in practice.
  
\end{abstract}

%==========================================================================
\section{Introduction}
The Lightning Network is a payment channel network using source-based onion routing to deliver payments from senders to recipients.
A necessary condition for a single onion package to be delivered successfully is that the onion follows a path with sufficient liquidity.
In this context sufficient liquidity does not just mean that the publicly known channel capacities of the channels on the path between sender and recipient are larger than the payment amount. Rather, every node along the path has to own enough of the channel capacity as their local balance to be able to forward the amount to the next hop.
As broadcasting the balance values would hinder the scalability of the Lightning Network they are generally kept private and thus unknown to other nodes.
Currently the sender node mitigates this uncertainty by entering a trial-and-error loop for delivering payments.
However, past experiments have demonstrated that payments are often failing, in particular when the amounts to be delivered are increasing\cite{DBLP:journals/corr/abs-1911-09432,DBLP:journals/corr/abs-2006-14358,lange2021impact, pickhardt2021security}.

Current implementations largely find candidate paths for the trial-and-error-loop by solving
shortest path problems or generalizations like $k$-shortest paths on the
weighted channel graph, where the weights correspond to the routing fees charged by nodes forwarding a payment along a channel.\footnote{We also observe a combination of the fee function with features like a penalty for longer CLTV values, prior experiences of using the channel and a bias against smaller channels.}
That approach tries to find the cheapest payment path for the sender but does not systematically factor in success probabilities. As a consequence, the payment loop might try a large number of cheapest but unreliable paths before timing out instead of using slightly more expensive but vastly more reliable paths. It also does not produce an optimal split of the payment into multiple paths.

Software implementations of the Lightning Network protocol have mainly focused on three strategies for handling the uncertainty of sending a payment.
\begin{enumerate}
\item Incentivizing the path finding algorithm to favor larger channels.\footnote{\url{https://lists.ozlabs.org/pipermail/c-lightning/2021-May/000203.html}}%clightning dev
\item Ad-hoc splitting of large payment amounts into smaller ones after failed attempts using a technique called multi-part payments (MPP).%\footnote{\url{https://www.coindesk.com/multi-part-payments-could-bring-bigger-bitcoin-sums-to-lightning-network}}
\item Using provenance scores of nodes and channels and other data collected during operation of a node to estimate which nodes and channels might be reliable.
\end{enumerate}

In this work we are developing a general technique that is achieving the effects of these rather ad-hoc techniques in a systematic fashion.
This text largely builds upon and extends prior research which pointed out that
the uncertainty of channel balance values can be mathematically modeled to
arrive at a probabilistic path finding scheme\cite{pickhardt2021security}. While
this earlier work demonstrated a significant reduction in failed attempts while
delivering payments it still mostly kept the perspective of a single path finding
problem.

It has long been a folklore view that delivering payments on the Lightning Network can be modeled as a network flow problem.
In what follows we show that the discovery of the most likely multi-path payment is equivalent to solving a (generalized) min-cost flow problem in which the negative logarithms of channel success probabilities are considered as the cost of using a channel.
The channel success probabilities are priors that have to be estimated, through sampling for example.
Under the assumption of an independent uniform balance distribution on each channel as the prior, finding the most probable multi-part payment for a given amount from a sender to a recipient can be usefully modeled as solving an integer min-cost flow problem with a separable convex cost function.
While in general min-cost flow problems are complex
optimization problems, the above mentioned subclass is known to have polynomial
time solutions with a runtime complexity of $O(m \cdot \log(U) \cdot S(n,m))$
where $n$ is the number of nodes, $m$ is the number of edges on the network, $U$
is the amount to be delivered, and $S(m,n)$ is the time it takes to obtain a
solution to the single source shortest path problem\cite{Minoux1986,ahuja1993network}. This
is typically done using Dijkstra's Algorithm in time $O(m+n) \cdot \log(n)$, so
that we arrive at a total runtime of $O(\log(U)\cdot(m^2+mn)\cdot\log(n))$.

If the balance values were known, the decision if a payment between two nodes can be conducted could be arrived at by finding a max-flow / min-cut and comparing it to the amount that is to be delivered.
Given the uncertainty of channel balances, the decision problem is much harder to solve and still involves a trial-and-error loop.
We introduce a practical round-based algorithm that can either quickly deliver the payment or decide with a high probability that the min-cut is lower than the payment amount that is supposed to be delivered between sender and receiver.
It starts by creating the most likely MPP split as the solution of the min-cost flow problem and sending out the partial payments.
The algorithm reduces its uncertainty of the balance values by learning from the failures and successes.
This is done by updating the prior probabilities after the failing onions have returned.
Finally it creates another candidate MPP for the residual amount by solving a min-cost flow on the updated graph.

\section{Payments as Integer Flows}
\label{flows}
Let $G = (V,E)$ be a directed graph and $u: E \longrightarrow \mathbb{N}$ a function assigning capacity values to all edges in the graph.
For every node $v\in V$ let $b_v \in \mathbb{Z}$ denote its \emph{excess, supply or demand}.  Typically $b_v$ will be $0$ except for the source node $s$ (with supply $b_v>0$) and the destination $d$ (with demand $b_v<0$). We call a function $f: E \longrightarrow \mathbb{N}_{0}$ a \emph{flow} if the following conditions hold:
\begin{enumerate}
\item \textbf{capacity constraint}: For every edge $e \in E$ we have: $$0 \leq f(e) =: f_{e} \leq u_{e} := u(e).$$
\item \textbf{flow conservation}: For every node $i\in V$ we have: $$\sum_{(i,j)\in E} f_{ij} - \sum_{(j,i)\in E} f_{ji} = b_i.$$
\end{enumerate}

\subsection{Flows on a Known Balance Graph}
Assuming the balance values of every payment channel of the Lightning Network were publicly known, one could create a flow network called \emph{balance graph} as follows:
For any given payment channel between two nodes $i$ and $j$ with capacity $u$ we know that the balance $u_j$ of node $j$ plus the balance $u_i$ of node $i$ must be equal to the capacity $u$.
On the balance graph we add two directed edges for the payment channel between the nodes $i,j$.
First we add the edge $(i,j)$ with a capacity of $u(i,j) = u_i$ and then we add another edge in the opposing direction $j,i$ with a capacity of $u(j,i)=u_j$.\footnote{In reality, channel reserves would need to be accounted for by lowering the capacities to the spendable balances.}

Observe that in this balance graph, a set of payment paths from $s$ to $t$ determines a flow simply by summing up the amount sent through any edge.\footnote{We abstract from the fact that the amount sent through a payment path diminishes along the path by the fees collected; we assume the total fees are included until the end. See Section~\ref{generalized} for a discussion of the impact of this relaxation.}
In fact, the converse is also true: It is easy to see that any flow can be decomposed into paths from $s$ to $t$ and cycles in linear time~\cite[p.~79ff]{ahuja1993network}.
Since cycles do not change supply or demand at any node, we can safely ignore them.
In this way, we can represent any MPP split as a flow and vice versa.
We emphasize that the paths of the decomposed flow that lead to the MPP split do not need to be disjoint.

Given the balance values, we could decide the maximum possible amount that can be sent for any given node pair $s,t \in V$ by computing the minimal $s$--$t$-cut.
Using, say, the Ford-Fulkerson algorithm\cite{ford_fulkerson_1956} one could compute a max-flow and disect it into a series of paths.
These paths could then be used to construct several onions for a multi-part payment.

\begin{figure}[htpb]
  \center
  \includegraphics[width=0.45\textwidth]{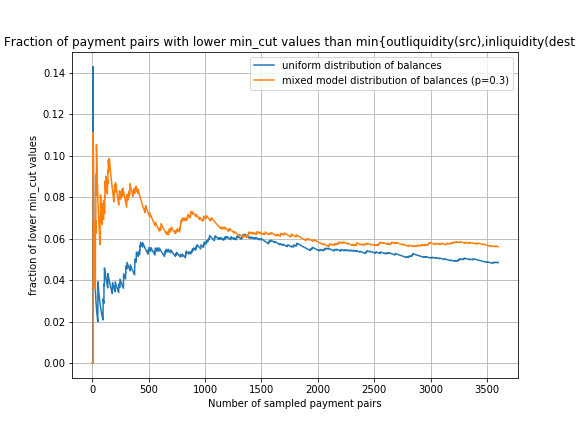}
  \caption{Showing the percentage of payment pairs where the maximal payable amount is actually lower than the upper bound given by the local balance known to both sender and recipient.}
  \label{fig:mincut}
\end{figure}

In a preliminary test we use two different prior distributions to generate two static balance graphs and check on each how often the min-cut between arbitrarily chosen pairs of sender and receiver is actually determined by the minimum of the local outbound capacity of the sender and the local inbound capacity of the recipient.
In Figure~\ref{fig:mincut} we can see that for both of these distributions --- which have been observed on the entire Lightning Network and an active core subset, respectively --- in only about $5\%$ of the payment pairs is the max-flow smaller than the maximum amount locally sendable by the payer and the maximum amount locally receivable by the target node.

The fact that with publicly known balance values in $19$ out of $20$ payment pairs the amount that can be delivered is as high as the local limits of the payment pair is in stark contradiction to the currently observed and reported \cite{DBLP:journals/corr/abs-1911-09432,DBLP:journals/corr/abs-2006-14358,lange2021impact, pickhardt2021security} success rates. In fact, these are declining heavily with amounts larger than $100,000$ satoshi (1 BTC = 100,000,000 sat) for which the delivery should be almost always possible.
We conjecture that this is due to the fact that in reality, the balance values are not publicly known.
This forces us to take the total channel capacities $u$ as capacities on our flow network in both directions.
Note that finding a max-flow on this network is not sufficient for deciding if a payment can be made. % removed footnote as it is actually a bit more subtle in reality https://twitter.com/renepickhardt/status/1412383409578590209
% TODO: do we want to include one sentence about the push relable algorithm by Elias Rohrer and explain why it should not work to well? Saturating edges.
However, we can generalize the flow model to the case of uncertain balance values and will return to this question in Section~\ref{rounds}.

\subsection{Uncertainty Networks}
Earlier research~\cite{pickhardt2021security} has introduced a mathematical framework for handling uncertain balance values in payment channels with the goal of making path finding decisions that maximize the success probability of payment paths.
We recall that --- given a prior belief about the balance uncertainty via a probability distribution $P_e$ and a random variable $X_e$ ---  the channel success probability for a payment of size $f_e$ for a channel $e$ is expressed as $P_e(X_e\geq f_e)$.
Whereas in~\cite{pickhardt2021security} the goal was maximizing the path success probabilities, here we aim to maximize the success probability for the entire flow. Assuming the channel balances to be independently distributed, this combined success probability is simply the product of all channel success probabilities:

\[
P(f)=\prod_{e\in E}P(X_{e} \geq f_e)
\]

Any flow $f$ that maximizes the success probability $P(f)$ is also minimizing $-\log\left(P(f)\right)$ and vice versa.
%The following sentence is a mathematical beauty and it would be really sad to have it removed (: 
Using the fact that the logarithm is a group homomorphism from the multiplicative group of positive real numbers to the additive group of real numbers we can write:
\[
-\log\left(\prod_{e\in E}P_e(X_{e} \geq f_e)\right) = \sum_{e\in E}-\log\left(P_e(X_e \geq f_e)\right)
\]
The right hand side of the equation has the form of a separable cost function $C$ for the flow $f$ from the theory of min-cost flows:
\[
C(f) := \sum_{e\in E}-\log\left(P_e(X_e \geq f_e)\right)
\]
Given fixed probability distributions for the channel balances, finding the most likely flow is therefore equivalent to finding a flow of minimum cost on this \emph{uncertainty network}. Any such flow can then be disected into the most likely multi-path payment as in the previous section.

In general, finding optimal solutions to the min-cost flow problem with non-linear costs is NP-hard~\cite{guisewite1990minimum}. Fortunately, in the special case of integer flows and capacities together with a separable convex cost function a polynomial algorithm has been introduced by~\cite{Minoux1986}.
Since our flows are integer-valued and the cost function is separable we need to understand when a cost function arising from channel success probabilities is convex in order to be able to apply such an algorithm.
Because the cost function $C$ is separable we can test convexity independently for any given channel $e$ and the resulting cost function $c_e(f_e):=-\log\left(P_e(X_e\geq f_e)\right)$
After simplifying by assuming a flow value $x:=f(e)$ and defining $p(x):=P_e(X_x\geq x)$ we get:
\[
c_e(x)=-\log(p(x))
\]
Assuming this function is twice differentiable, it is convex iff its second derivative is be nonnegative on the entire domain.
The first derivative is:
\[
c_e'(x) = -\frac{p(x)'}{p(x)}
\]
and the second derivative is:
\[
c_e''(x) = \frac{(p'(x))^2-p(x)p''(x)}{{p^2(x)}} \geq 0
\]
In particular we see that the negative log probabilities result in a convex cost function iff the following inequality holds:
\[
(p'(x))^2 \geq p(x)p''(x)
\]
In the uniform case $p(x)=\frac{u+1-x}{u+1}$ (cf.~\cite{pickhardt2021security}) we have $p'(x)=\frac{-1}{u+1}$ and $p''(x)=0$ demonstrating that the resulting cost function $c_e(x)$ is convex on its entire domain.
This indicates that the polynomial algorithm can be used to find the flow that minimizes:
\[
C(f)=\sum_{e\in E}-\log\left(\frac{u_{e}+1-f_e}{u_{e}+1}\right)
\]
Thus the most probable multi-part split for delivering the amount $U$ can be found by solving a min-cost flow problem and disecting the flow into paths.

We do not explicitly handle fees in this model, but observe that if we can find a flow that includes an upper bound to the total fees, the real payment success probability will be at least as high as the one predicted by this model, since the transported amount is only falling along the paths and our probability function is monotonic.

%In~\cite{pickhardt2021security} it has been shown that the channel success probability for a payment of size $f_e$ for a channel $e=(i,j)$ of capacity $u$ between $i$ and $j$ can reasonably be modelled as uniformly distributed.
%Thus the channel success probability $P(X_{e} \geq f_e)$ can be expressed in terms of the assigned flow as $ \frac{u+1-f_e}{u+1}$.
%In order to find the best multi-part split when delivering the amount $U$ from $s$ to $d$ we are searching for a flow $f$ with the appropriate supplies and demands that maximizes $P(f)$.
%Maximizing a probability is equivalent to minimizing its negated logarithm.
%Thus we can also find a flow that minimizes

\subsection{Maximizing Success Probabilities vs Minimizing Fees}
\label{fees}
The current routing fee function on the Lightning network is a separable cost function depending only on the flow across each channel.
However, it is easy to see that the function is not convex at the transition between flow $0$ (cost $0$) and flow $1$ (cost base fee plus unit flow cost), whenever the base fee is larger than the proportional unit flow cost. In fact, a cost function of this form is often called a \emph{concave fixed-charge cost} in the literature. Unfortunately, finding the flow that minimizes a cost function of this form is a strongly NP-hard problem as shown in~\cite{guisewite1990minimum} by reduction from 3-SAT to a min-cost flow problem with only fixed-charge costs.

On the other hand, if the Lightning Network community were to drop the base fee, the separable cost function would become linear in the flow value of each arc.
Finding an MPP split that minimizes routing fees could easily be done by solving the linear min-cost flow problem using any of a number of algorithms~\cite{ahuja1993network}.
However, we note that minimizing the routing fees alone tends to
saturate the full capacity of cheap channels. Such paths are highly improbable to succeed since they can only do so when the entire balance is on the right side of the channel (even ignoring channel reserves). In our opinion that makes optimizing purely for fees a poor choice in general. On the other hand, only maximizing the success probability might allow routing node operators and liquidity providers to game the algorithm and extract unlimited fees.

So in practice, it should be our goal to both minimize fees and maximize success probabilities.
Naturally, these goals can be contradictory, since node operators
can and will choose fees freely. Two ways of expressing this goal might be
\begin{enumerate}
\item to minimize fees with a side constraint of achieving a certain minimal probability bound, or
\item to maximize success probability with a side constraint of staying below a certain maximal fee bound.
\end{enumerate}

%TODO: removed the "it is easy to see part. I don't see it easily"
Unfortunately, both of these problems are weakly NP-hard via the following argument:
First, observe that adding unit capacities to the problems and looking for a flow of size 1 makes them instances of the so-called constrained shortest-path problem.
Then,~\cite[p.~798]{ahuja1993network} shows that this subclass of constrained min-cost flow problems is already NP-hard by reduction from the Knapsack problem.

Fortunately, the picture is not quite as bleak as it looks on first sight.
First, the reduction only implies weak NP-hardness, meaning that we could find a polynomial algorithm whenever $U=O(n^k)$ for some $k$.
Strictly speaking, the flow size $U$ is always bounded by a constant in our applications, since the total number of bitcoin is limited. Looking into the theory of Lagrangian relaxation~\cite[p.~598ff]{ahuja1993network} methods, however, gives us immediate practical results instead of just hope.

In fact, the two cases enumerated above collapse into one when we try to find bounds for them by applying a simple one-dimensional Lagrange multiplier, that is we try to minimize:
\[
\sum_{e\in E}-\log(\frac{c_{e}+1-f_e}{c_{e}+1})+\mu\cdot f_e\cdot fee(e)
\]
the linear combination of both costs, with a suitable multiplicative constant
$\mu$.
By calculating this combined min-cost flow (note that the linear combination of
the two cost functions remains convex), not only do we get a feasible flow of size $U$, but because of the \emph{Lagrangian bounding principle}~\cite[p. 605f]{ahuja1993network}, whatever total fee $x$ and success probability $p$ we achieve, we are guaranteed that this combination is optimal on both sides.
That is, there is no solution with cheaper total fees than $x$ and at least probability $p$, and there is no solution with higher success probability than $p$ that still only costs at most $x$.
So in case we are not satisfied with the parameters of the solution we got, finding an adequate solution is reduced to either increasing $\mu$ and getting a potentially cheaper, but less probable solution, or decreasing $\mu$ and receiving a more reliable but also more expensive solution.

\subsection{Generalized Flows with Losses}
\label{generalized}
So far we have ignored the fact that every node on every path takes some part of the payment as a fee, which means that the total amount of flow gets smaller towards the target.
This observation is best described by a slightly more general model called \emph{generalized (minimum cost) flows with gains and losses}. In this formulation, the flow conservation condition for any node $i\in V$ is changed to $$\sum_{(i,j)\in E} f_{ij} - \sum_{(j,i)\in E} \gamma_{ji}f_{ji} = b_i.$$
Thus, when we send 1 unit of flow along an edge $(i,j)$, $\gamma_{ij}$ units of flow arrive at node $j$. The edge multipliers $\gamma_e$ are positive rational numbers\footnote{In the Lightning Network, these correspond to the proportional part of the fee. Again, including the base fee makes solving the problem infeasible, which is why we propose abolishing it.} and the edges are called \emph{gainy} ($\gamma_e > 1$) or \emph{lossy} ($\gamma_e < 1$) accordingly.
Notice that until now, this formulation still depends on us knowing the exact supply/demand amounts at the source and destination nodes. This is especially troubling here, because we cannot just use an upper bound to the supply as before: there might not be a solution that uses the corresponding exact amount of fees. Therefore we introduce a cost-free high-capacity gainy (say, $\gamma_s=2$) self-loop at the source node and set the source excess to zero. This allows for introduction of an arbitrary amount of flow at the source. Then, we aim to minimize the convex cost function as before under the remaining flow and capacity constraints.

The generalized flow problem is clearly a proper generalization of the min-cost flow problem outlined above. Unfortunately, it also appears to be harder to solve. The algorithm that we implemented for the min-cost flow problem does not seem to carry over to this more general problem. We did find pointers to some approaches~\cite{tsengbertsekas} that might be worth exploring. So far, we have been reluctant to invest too much effort in this direction, because in our application, the fees are generally expected to be a small fraction of the total flow.  Thus it is doubtful if the greater computational effort will be worth the slightly more favorable probability/fee result.

\section{Payment Algorithm}
\label{rounds}
Once we are able to efficiently compute minimum cost flows optimizing either for success probabilities, fees or both using the Lagrange relaxation we naturally arrive at a round-based payment algorithm that can be used by a node $s$ that wishes to send an amount $U$ to a destination $d$.
For now, we assume a Lagrange-style combination between channel success probabilities and the linear fee rate function, as we believe this achieves the most useful results.

The round-based algorithm is initialized by the sending node $s$ in the following way:
It starts by creating a new uncertainty network model $N$ of the Lightning Network which encodes the initial uncertainty and information $s$ is gaining about the balance distribution on the network during the rounds.
%TODO: put back in? Thus the network will be updated during the process of delivering the payment during the following rounds.
In order to deliver the full amount, the node $s$ will have to solve a minimum cost flow problem, send out onions and update the uncertainty network based on the successes and failures of the onions in each of the rounds.
The uncertainty network $N$ consists of all the nodes that are on the same connected component as $s$ on the channel graph.
The edges of $N$ correspond to payment channels on the Lightning Network.
If $s$ has no further knowledge about the channels, a directed arc for both directions of each payment channel is added to the uncertainty network.
The capacities of the edges are set to the capacities of the payment channel (possibly deducting channel reserves).

Notice for example that in the local channels of $s$ the balance values are known and there is no uncertainty.
Thus the capacity for those channels is set to the local balance value $u$ as this is the most that can currently be sent on those channels.
The probability is set to $1 = P(X\geq a | X=u)$ for any amount $a$ between $0$ and $u$.
This results in a negative log probability of $0$ and thus makes it very cheap for the minimum cost flow computation to utilize the liquidity in this channel. In particular since the node $s$ also does not have to pay any fees to itself.

Similarly, the receiving destination node $d$ could tell the sending node about the state of its local channels and this knowledge could also be incorporated into the graph by creating edges with 0 log probabilities and decreased capacities.\footnote{Communicating this information in invoices is currently not part of the protocol but routing hints in the invoices might be extended to encode such information.} In Figure~\ref{fig:mincut} we have demonstrated that for about $95\%$ of all payment pairs the amount that can be delivered through the network is actually limited by the local outbound capacity of the sender and the local inbound capacity of the receiving node which yields another motivation for this information to be shared.

After the setup phase the round-based phase starts. Here the algorithm iterates over the following steps until either the entire amount is delivered or the minimum cost flows become too unlikely or cannot be computed for the residual amount, which means the minimum $s$--$d$-cut has been discovered.\footnote{In such cases it seems reasonable that $s$ open a new payment channel with $d$ for at least the remainder amount.}
The round starts with $s$ computing a minimum cost flow for the amount $U$ to $d$ following the optimization goal.
The flow is then decomposed into paths and cycles. Note that cycles cannot appear in our application as long as we do not allow negative cost edges. If they appeared, they would indicate profitable rebalancing options that may or may not be reachable for $s$. Since negative fees are not allowed in the BOLT standards, we can safely ignore cycles for now.

The node then conducts a multi-part payment by concurrently sending out onions along the resulting paths.
In practice one has to chose a decomposition of the flow into paths that does not create more HTLCs on a remote channel than the HTLC limit encoded into the protocol permits.
This engineering challenge as well as others like the question of channel reserves are ignored here for simplicity of the presentation.
Despite the fact that the most likely flow was used, some of the onions will not reach the target in most cases. So the sending node gathers information from the error codes of failed attempts as previously described in~\cite{tikhomirov2020probing} as well as information from the paths that have not returned an error to update the probabilities as described in~\cite{pickhardt2021security}.
This step decreases the uncertainty of the channel balances and is crucial for the improvement and different results in the next round which is why we explicitly explain how the knowledge is updated in several cases.
\begin{enumerate}
\item If an onion with the amount $h$ has not been returned, we assume it has arrived at the destination. Thus all channels across the path have now locked some liquidity into an HTLC.\@ In our uncertainty network we thus reduce the capacity $u$ of each involved channel by the amount $h$ that was sent along that channel on that path. This changes our future success probabilities for the amount $a$ to be $P(X\geq a+h|X\geq h)$ which corresponds to a change from $\frac{u+1-a}{u+1}$ to $\frac{(u-h)+1-a}{(u-h)+1}$ in the uniform case. For any value of $a$ and positive $h$ the second fraction is smaller that the first one. This leads to lower probabilities which in turn yields higher costs to use those channels again in follow up rounds.
\item If an onion of size $h$ fails we learn the following new information:
  \begin{enumerate}
  \item On every channel up to the failed channel there has been sufficient liquidity to forward the amount $h$. In future rounds we can use the conditional probability $P(X\geq a|X\geq h)$. For $a<=h$ this term will be $1$, resulting in log costs of $0$. For $a>h$ the conditional probability is uniform again. It is important to notice that the conditional probability will again lead to a convex cost function.
  \item Assuming the failure code is related to liquidity issues for the failed channel we know that the amount $h$ was not available. Thus we are able to reduce the capacity to $h-1$ and change the probability for the future rounds from $\frac{u+1-a}{u+1}$ to $\frac{h-a}{h}$ in the uniform case, or $P(X\geq a|X < h)$ in general. This probability decrease will result in higher costs for utilizing this channel.
  \item If on the other hand the failure code is related to an issue with the next hop, for example a downtime, the node can update its local view by temporarily removing the failed node with all its channels.
  \end{enumerate}
\end{enumerate}

Note that both successful and failing onions would also allow us to update the knowledge about the balances in the reverse direction.
Once all the knowledge from the partial attempts of the round has been learned --- which is reflected by the update of the probabilities and cost functions --- the algorithm enters the next round.
First, the sum of all failed onions is computed as the residual amount that still needs to be delivered.
We enter the next round with that amount on the updated uncertainty network by computing a new minimum cost flow to generate the next optimal MPP split.

\section{Example}
\begin{figure}[htpb]
  \center
  \includegraphics[width=0.45\textwidth]{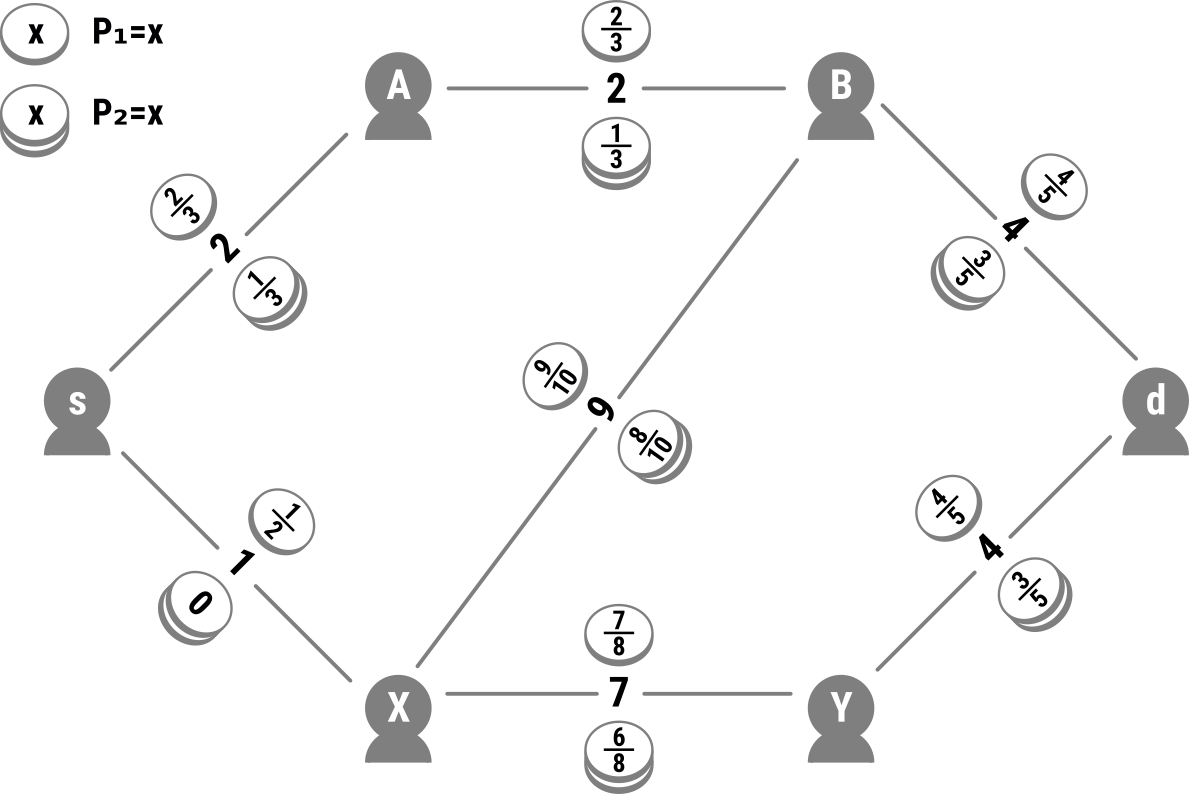}
  \caption{Example Channel Graph on which the $2$-flow with maximal probability is not an extension of the $1$-flow from  $s$ to $d$ with maximal probability. The success probabilities for sending $i$ sat are depicted as $p_i$. The channels capacities are the bold numbers along the edges.}
  \label{fig:counterexample}
\end{figure}

Let us go through an example that illustrates why finding a min-cost flow is necessary instead of just adding optimal paths.
It also serves to better understand the round-based algorithm.
In order to simplify the example, we ignore fees and channel reserves and optimize purely for probability here.
We also hide the nonintuitive negative log probabilities by writing down the corresponding max-probable flow with probabilities as edge weights.
For the channel graph depicted in Figure~\ref{fig:counterexample},
assuming uniform probability distributions we compute the following probabilities for the $1$-flows (paths delivering 1 sat):
\[
\begin{aligned}
p([s,A,B,d]) = & \frac{2}{3}\cdot\frac{2}{3}\cdot\frac{4}{5}=\frac{16}{45} &= 0.35\overline{5} \\
p([s,X,Y,d]) = & \frac{1}{2}\cdot\frac{7}{8}\cdot\frac{4}{5}=\frac{28}{80} &= 0.35 \\
p([s,X,B,d]) = & \frac{1}{2}\cdot\frac{9}{10}\cdot\frac{4}{5}=\frac{36}{100} &= 0.36 \\
\end{aligned}
\]

This indicates that $s,X,B,d$ is the minimum cost $1$-flow.
The $(s,X)$ arc is obviously saturated so that a $2$-flow extending the $1$-flow would have to go via the $(s,A)$ arc.
One can easily compute the probability of the resulting $2$-flow $f^2$ if the next sat is also using the $(B,d)$ channel and being merged with the min-cost $1$-flow as
\[
p(f^2) = \frac{1}{2}\cdot \frac{9}{10}\cdot \frac{2}{3}\cdot \frac{2}{3}\cdot \frac{3}{5} = \frac{3}{25}=0.12
\]

However if we look at the $2$-flow $g^2$ that sends $1$ sat along $s,A,B,d$ and $1$ along $s,X,Y,d$ we get $p(g^2)=\frac{16}{45}\cdot\frac{28}{80} = 0.124\overline{4}$
which is also the min-cost $2$-flow in this graph.
This example shows that finding a min-cost flow cannot in general be done  by computing the most likely path for a single sat and extending it with the next most likely $1$-sat-paths.\footnote{This simple $+1$-algorithm could actually be rescued so that it would be able to compute the min-cost flow. However both versions would also be quite slow as they would be linear in the amount that was to be sent --- which is exponential in the input size}

\begin{figure}[htpb]
  \center
  \includegraphics[width=0.45\textwidth]{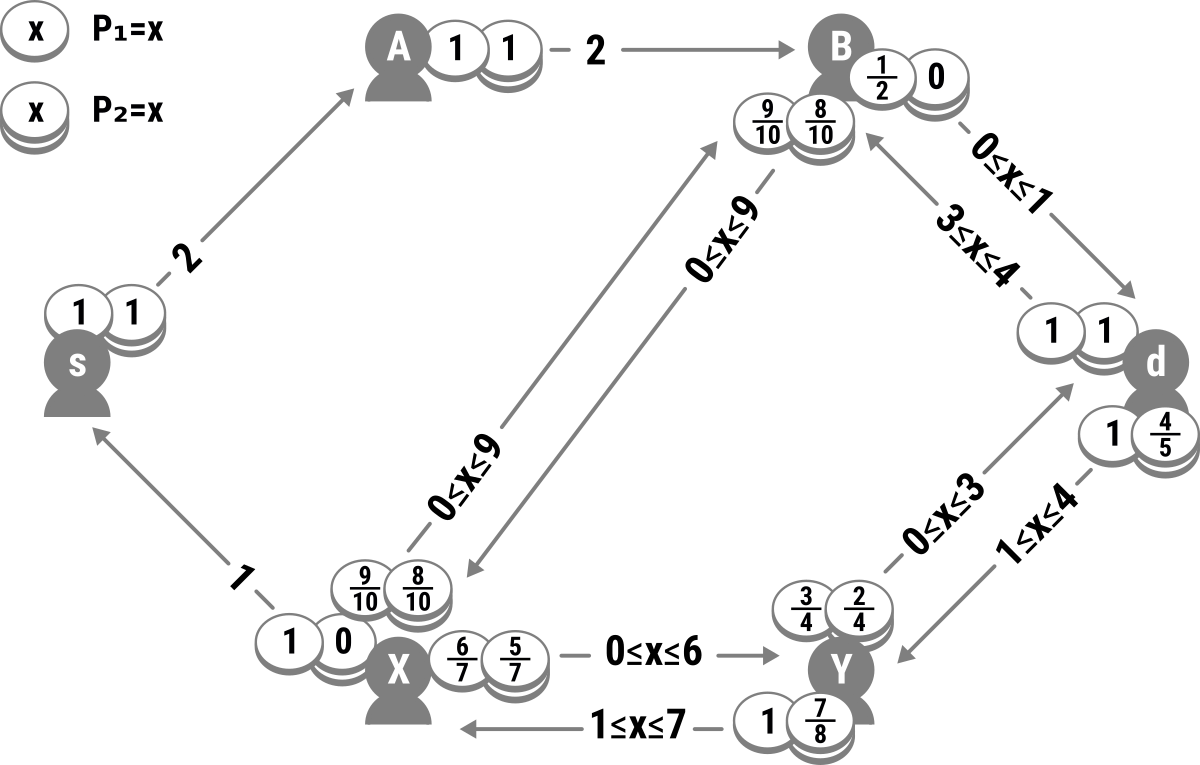}
\caption{The uncertainty network of $s$ after HTLCs of value $1$ are locked in along the path $s,X,Y,d$ and a 2-sat onion along $s,A,B,d$ failed because of missing liquidity on the $B,d$ channel. Saturated edges are removed. The black labels on the edges depict the uncertainty range of the balance or a single number if the balance is known to $s$. The $p_i$ express the success probability for sending a further $i$ sat along an edge given the updated knowledge. }
  %  \caption{The arc $(s,X)$ has been removed because it was fully saturated in round $1$. The arcs $(s,A)$ and $(A,B)$ are marked with cost $0$ or probability $1$ as they could deliver $2$ units in the first round. The arc $(B,d)$ is effectively reduced to capacity $1$ as it could not deliver $2$ sats in the first round. Finally the capacities of arc $(X,Y)$ and $(Y,d)$ are both reduced by $1$ as $1$ unit is currently locked into an HTLC.}
  \label{fig:round2}
\end{figure}

Extending our example and assuming we want to send a total of 3 sat we start again by computing the min-cost flow $f^3$ which can be disected into two paths $l_1= s,X,Y,d$ with an amount of $1$ and another path $l_2=s,A,B,d$ with an amount of $2$.
After sending out the onions we might have learned that the onion along $l_1$ has been successful, but the one with $2$ sat along $l_2$ has failed because $B$ did not have enough liquidity to forward the onion to $d$ on the $(B,d)$ channel.
For the second phase of the algorithm we now compute the min-cost flow on a graph where we know that we can deliver 2 sat with perfect certainty to $B$.
This updated uncertainty network is depicted in Figure~\ref{fig:round2}

The flow that sends the full residual amount of $2$ sat along $s,A,B,X,Y,d$ has a probability of $\frac{8}{10}\cdot\frac{5}{7}\cdot\frac{1}{2}=0.286...$ while the flow that sends $1$ sat along the $B,d$ channel has a probability of $\frac{1}{2}\cdot\frac{9}{10}\cdot\frac{6}{7}\cdot\frac{3}{4}=0.289...$ telling us that for the residual amount we should make another split payment sending one sat along each of the paths $q_1=s,A,B,d$ and $q_2=s,A,B,X,Y,d$.

Finally, if the path $q_1$ locks in and $q_2$ returns an error at any of the channels $(B,X),(X,Y)$ or $(Y,d)$ we would know that we cannot deliver the full payment as the min-cut in the network on the balance graph had the value $2$.
If however $q_2$ and $q_1$ both lock in we have successfully delivered the payment.
If both $q_1$ and $q_2$ return an error we know the min-cut between $s$ and $d$ on the balance graph was $1$ (as that had been locked-in in the first round and no further HTLCs have locked in).
Finally if $q_1$ returns an error and $q_2$ locks in we will have to enter the third round.
In the third round there is only 1 sat to be delivered on a single possible path $q_2$ which, given our knowledge, has a success probability of $\frac{8}{9}\cdot\frac{5}{6}\cdot\frac{2}{3}=0.494\ldots$\footnote{the numerical similarity to the twitter account mentioned in the acknowledgements is completely coincidental as we used that graph even before we had the discussion with the individual.}.

\section{Anecdotal Lab Report}
\textbf{Disclaimer:} We stress that this document is a preprint. In particular this section cannot be considered a proper evaluation. For that we would also have to test the algorithms on the Lightning Network mainnet.
In addition to the fundamental complications arising through the base fee\footnote{which should be easily avoidable at the moment by incorporating some buffer in the fee size, as long as it remains lightly used}, such a real-world test requires overcoming several engineering challenges that we did not have the time to address yet:

A practical implementation needs to automatically answer questions that might arise in case of hanging HTLCs, if channels become inactive, or amounts reach HTLC limits of channels, for example.
Also the implementation of the min-cost flow solver would have to be engineered to have a much faster runtime than our experimental version so that it would actually be feasible to use it on the real Lightning Network.
However we felt the need to share some preliminary research progress after some test results from our simulated setting indicated an improvement of reliability of several orders of magnitude over the currently reported statistics for payment delivery.
In particular, an anecdotal lab report --- while highly unusual --- seemed appropriate for this particular situation in order to inform the developer community as early as possible about the potential need for a protocol upgrade removing the base fee.

\subsection{Simulation}
Since the computation with base fees is not feasible\footnote{and the fees are also currently low enough to not impact the results too much}, we have ignored the base fees in all our computations.
In fact, we started experimenting by not even optimizing for low fees at all but just for high probabilities.

We took a recent snapshot from the channel graph of the Lightning Network that was observed via the gossip protocol on one of our long running mainnet nodes.
We then created a static balance graph instantiation of the simulated network by splitting the channel capacity into two balance values uniformly at random independently for each channel.

We created a Python-based min-cost solver following the algorithm described in~\cite[p.~556ff]{ahuja1993network} and a Scala version later on that turned out to be faster by a factor of about $3$--$4$.
While implementing this algorithm we made some mistakes early on that accidentally led to the discovery of a heuristic that, on our snapshot Lightning Network graph, reliably produced results with less than 1\% deviation from the optimal cost in less than 1\% of the runtime.
Because the exact algorithm takes more than 30 minutes even in the faster Scala implementation, the following results have been mostly obtained with this heuristic, which typically takes about 6 seconds to run in Scala. So it is notable that the optimal results would be even better, although negligibly so.

We picked a medium-sized Lightning node that got randomly assigned a local balance of $0.42$ BTC and tried to send $0.4$ BTC to another node $3$ hops distant using the round-based algorithm described in Section~\ref{rounds}.
The remote node had a total capacity of roughly $1.5$ BTC and more than $0.4$ BTC inbound liquidity. 
Assuming no routing hints from the recipient we started the first round computing a flow that was disected into several hundred small paths.
Sending them out we where able to deliver almost $75\%$ of the amount that we wanted to deliver at once.
We updated the graph with the insights from the successes and failures and started the second round for the residual amount of roughly $0.1$ BTC.
In this round the min-cost solver on the graph with less uncertainty suggested a split of about $100$ paths.
After sending out the payments we observed that the residual amount was only about $0.009$ Bitcoin.
In the third round, again on the updated graph, the min-cost solver suggested to send about 15 concurrent payments, of which all but one where successful.
We entered the fourth and final round with an amount of $30,000$ sat ($0.0003$ BTC) remaining.
Owing to the learned data, the heuristic of our min-cost solver sent the full amount on a single path with $8$ hops, because it had already gained enough certainty for all but one channel (of size 1 BTC) along that path that it could forward $30,000$ sat.
Thus with a $99.97\%$ probability the $30,000$ sat path settled on the selected $8$ hops path and the payment was delivered in full.
While sending out all the onions we tracked the total routing fees to be $814$ sat.

We also repeated the experiment with reduced initial uncertainty by assuming the recipient node had initially communicated to the sender on which channels it could receive what amounts in the invoice.
In this case and on the same graph the algorithm delivered the final payment in the third round already.

We repeated the experiment a couple of times with different amounts and different instantiations of the simulated balance graph, resulting in similar results with every run. We therefore believe it reasonable to expect that we would see very similar results on the actual Lightning Network even though the unknown balance graph of the mainnet is constantly changing (potentially making some of our learned knowledge invalid).

We also used the above combination of balance graph, source and destination node for some experiments with the Lagrangian relaxation. This time, we allowed multiple parallel channels between nodes, as they are actually observed in the Lightning Network. Also, instead of going through the rounds of the payment algorithm, we just looked at the results of a single min-cost flow calculation, with a payment amount of 9.2 million sat (0.092 BTC). In this setting\footnote{without using any knowledge about the source's or the destinations' channels}, optimizing for reliability only ($\mu=0$) yields a probability of $P=0.16$, with total fees of 697 sat (excluding base fees). On the other hand, choosing $\mu=100$ means optimizing almost exclusively for fees. This brings the total fees down to $16$ sat, but, as expected from the arguments in Section~\ref{fees}, the success probability drops to $P=1.1\cdot10^{-11}$. Table~1 shows that when we lower $\mu$ through multiple orders of magnitude, the success probability increases drastically, while the fees are only rising moderately. Unfortunately, we increasingly observed numerical instabilities while decreasing $\mu$. This lead to our algorithm not terminating beyond $\mu=0.01$.

\begin{table}
\begin{center}
\begin{tabular}[h]{l|l|r}
  $\mu$ & P & fee(sat) \\
  \hline
  100  & $1.1\cdot10^{-11}$ & 16 \\
  10 & $2.3\cdot 10^{-5}$ & 16 \\
  1 & $0.0097$ & 18 \\
  0.1 &  0.044 & 24 \\
  0.01 & 0.056 & 28 \\
  0  & 0.16 & 697
\end{tabular}
\caption{Results of varying $\mu$ on a fixed payment pair}
\end{center}
\label{table:mu}
\end{table}

Again, all numbers are from the heuristic, but the exact algorithm performed nearly identically in all our samples, including the numerical instabilities.

\subsection{Source code}
We could not find any preexisting open source software implementing the solution of the integer minimum cost flow problem for arbitrary separable convex cost functions.
We therefore share the source code of the described algorithms and methods as well as the latex sources of this document with an open MIT license at: \url{https://github.com/renepickhardt/mpp-splitter}.
This repository consists of Scala- and Python-based example implementations of the exact min-cost flow algorithm described in~\cite[p.~556ff]{ahuja1993network} for a separable convex cost function.
It also includes a minimalistic simulation framework in Python to test the practicality of the round-based payment loop.
The Scala version includes example code to demonstrate the usage of the Lagrangian relaxation.

\section{Advanced Applications}

\subsection{Multiple Senders and/or Receivers}

Notice that our definition of a flow in Section~\ref{flows} allows for an arbitrary number of both sources and sinks, that is, nodes with non-zero excess. This means that while a min-cost flow calculation might be computationally expensive, it can result in an optimized flow for multiple payments and/or channel balancing efforts at the same time. With respect to the runtime of the algorithm we have implemented\footnote{which relies on single source shortest path calculations}, a more complex flow will take longer to optimize in practice, even though it will still respect the same worst-case runtime bounds.
We expect that entities like Lightning Service Providers (LSP) or trampoline routers, who need to find paths for many payments, will find this aspect helpful.
One could imagine a permanently running min-cost flow calculation loop that keeps learning about the network and sending out remainder amounts as in Section~\ref{rounds}, but can always add additional payments in the next round.
It bears mentioning that in such multi-purpose rounds, the minimum cost is always optimized globally, which could lead to some payments being cheaper at the expense of others.
This needs to be accounted for when, e.g., routing payments for multiple clients.

\subsection{Optimal Channel Rebalancing}
\label{rebalancing}
It is well-known in the community that routing nodes can contribute to the overall payment reliability in the Lightning Network by using various channel rebalancing techniques. A recent survey paper~\cite{papadis2020blockchain} in particular describes channel rebalancing via off-chain circular onions. This can happen proactively or lazily at routing time via a technique called just in time routing (JIT-routing)\cite{Pickhardt2019}. To our knowledge, rebalancing has so far only been considered one channel pair at a time.
We observe that a node $i$ might want to rebalance several channels at once by shifting excess balance from source channels to target channels where more liquidity would be demanded.
In the uncertainty network, we can then assign the supply for rebalancing to the channel partner nodes of the corresponding outbound supply channels and remove these edges from the graph.
Because we have to account for the fees on inbound channels, the construction is a little more involved for the channels that demand extra balance: For every incoming edge $(j,i)$ that demands a balance increase, we create a copy $(j,k)$
(with the same capacity and cost function) leading to a new node $k$ that is assigned the demand.
Finally, we can compute a multi-source-multi-sink min-cost flow in order to shift the liquidity and conduct a multi-channel rebalancing.
As rebalancing is rarely time critical, one might prefer a high value of $\mu$ in the min-cost flow computation that favors low fees over a high success rate.
A node might even decide to stop the rebalancing operation before all of the onions have been successfully delivered: it has just engaged into a cheap opportunity for rebalancing; if delivering the remaining amounts turns out to be too expensive in the next min-cost flow calculation, it might prefer to stay with this improved but not perfect balance according to its own rebalancing strategy.

At first sight, rebalancing seems most interesting for nodes that engage more in routing than sending or receiving payments.
However, we want to stress that for LSPs who conduct several payments per second, it might be very reasonable to combine the rebalancing and payment aspect, and suggest two ideas.
First, we recall the global uncertainty of a node is always decreasing while delivering payments, so learning this knowledge could help find opportunities for engaging in rebalancing operations.
Second, an LSP might aim to keep its channels balanced according to a certain strategy.
Instead of allowing itself to use all channels for making a payment, an LSP could restrict itself only to the channels where it has too much liquidity and exclude other channels from the computation.
This min-cost flow might be more expensive and less likely, but it might increase the chances for the node to forward payments; potentially earning a fee and increasing its overall reliability for other nodes might make such a trade-off worthwhile.

\section{Limitations}

\begin{enumerate}
\item As discussed in Section~\ref{fees}, the currently adopted base fees in the Lightning Network make computing a min-cost flow NP-hard whenever the cost function includes these fees. Thus we have inquired about the motivation for the inclusion of a base fee online and received a response from the developer who appears to be responsible for this decision.\footnote{Rusty Russell's answer to the question: Why was the base\_fee for the routing fee calculation of the Lightning Network included? \url{https://bitcoin.stackexchange.com/a/107341/81940}}
We are under the impression that the base fee was a rather ad-hoc and arbitrary choice and is not of significant importance to the Lightning Network protocol.
Even if it were too difficult to change the protocol we see clear incentives for node operators setting the base fee to zero:
Nodes who want to conduct path finding might ignore base fees in their path computation in the future and thus ignore channels with a non-zero base fee.
Furthermore, we note that our approach favors channels with large capacity, whose operators might therefore be able to demand higher fees.
We conjecture that this will give rise to a more balanced fee market, which should be in the interest particularly of node operators who have invested significant liquidity.
\item We made the crucial assumption that channel balances are uniformly distributed. While this was confirmed by prior research~\cite{pickhardt2021security}, the situation could evolve over time, making our priors less suitable.
If, say, we assumed a normal distribution, the negative log probabilities would not be convex on the entire domain and the min-cost flow problem might not be solvable in polynomial time.
However, we note for future research that there are practical solutions
for this, like limiting the domain by limiting the allowed channel capacity. Moreover, in~\cite{pickhardt2021security} it has also been shown in the single path case that adopting a rebalancing protocol which changed the prior to a normal distribution but computing paths with a uniform prior still performed well.
In fact, we conjecture that we cannot do better than assume uniformity unless we have knowledge of the parameters of the actual distribution.
\item While the min-cost flow problem admits a polynomial time solution in the case of a convex cost function (without base fees), it is still computationally intensive, with a runtime that is quadratic in the number of channels. Remember also that we might have to solve several of these problems per payment round in order to find a suitable Lagrange multiplier in the trade-off between reliability and fees.
Our prototype implementation is currently not optimized for speed; on the full Lightning Network graph, running times can easily reach 30 minutes and more, depending on the degree of precision. However, we have achieved preliminary experimental results on a promising heuristic that seems to very favorably trade reductions in runtime for only a slight deterioration from the optimal solution of the minimum cost flows.
Using heuristics like this, algorithmic methods including approximation and parallelization, as well as optimized code, we estimate that on the currently public Lightning Network it should be feasible to achieve sub-second runtimes on commodity hardware.
However, we feel that this research is still too early for publication.
\end{enumerate}

\section{Future work}
Beyond the optimization steps required for practical usability and extensive tests on the actual Lightning Network that we have
hinted at above, we see some additional directions for further research:

%\subsection{Heuristic of not carrying optimality through delta phases}
%\subsection{Rounding the amount}
%\subsection{Pruning the problem}

\begin{enumerate}
\item Recall from Section~\ref{fees} that when optimizing for both fees and reliability one has to find a suitable value for $\mu$. We hope that improving on the current Lightning Network user experience will be possible with some experimentation, user interface design, and drawing from the extensive literature on Lagrangian relaxation.

\item In the round-based payment algorithm described in \ref{rounds}, the optimal delay spent waiting for further responses after the first failed onion has returned and before entering the next round with updated knowledge remains an open question. In our experiments and simulation we have assumed waiting for the status of all onions to be resolved. We defer investigation of this question to practical experimentation but point out that it does not appear too critical since we can always incorporate information that arrives too late for one round in any later ones. In the extreme case, for every returned onion a new round could be entered.

\item It is conceptionally straight forward to extend the probabilities with provenance scores from operating nodes on the Lightning Network.
Instead of just looking at the channel balance distribution one could create a joint distribution from, e.g., estimating the nodes' uptime.

\item Large channels could become the equivalent to the autobahn and attract a lot of traffic. Given the limitations of concurrent payments on the Lightning Network, the need for congestion control mechanisms might arise.

\item The payment planning and execution algorithm accumulates knowledge about the actual channel success probability distributions. However, we know that in practice those distributions do not stay constant. In fact, they change with every payment that is conducted through a particular channel or via the rebalancing behavior of node operators.
Further research into the dynamics of the money flow through the Lightning Network will help estimate how long to rely on the knowledge gained from previous payment attempts. This knowledge can then be adequately discounted or forgotten in future rounds.

\item The promising idea of adding redundant liquidity during the path finding phase of MPP splits was introduced by~\cite{DBLP:journals/corr/abs-1910-01834}. It would be very interesting to study if we could find optimal redundant overpayments so that we can expect to finish within one round of sending out multi-part payments with high probability.

\item  In spite of the arguments given in Section~\ref{generalized} that led to prioritizing other aspects first, we still believe generalized flows to be an interesting future research direction, especially in light of multiple-source-multiple-sink applications (cf. Section~\ref{rebalancing}). In this setting, constructions like the self-loop we described might allow for more flexibility: One can specify some sources and destinations with exact excess requirements as well as leave others open in order to give the optimization more leeway.

 %TODO:  mention flow merging / splitting in case of several sink nodes as with trampolines and the potential trick with PTLCs that makes this possible? 

\end{enumerate}

\section{Acknowledgements}
This research was partially sponsored by the Norwegian University of Science and Technology (NTNU).
We want to thank David Karger and MIT open courseware for sharing a lecture series about min-cost flows\footnote{\url{https://www.youtube.com/playlist?list=PLaRKlIqjjguDXlnJWG2T7U52iHZl8Edrc}} that has been very useful for us in creating our reference implementation and understanding the theory of min-cost flows.
We are grateful to Ravindra Ahuja, Thomas Magnanti and James Orlin for their exceptional textbook~\cite{ahuja1993network} which contains so much of the knowledge we have been adapting to our use case here.
We also thank Twitter user \texttt{@four9four} for discussing our preliminary results after the Lightning hacksprint in April 2021.
Thanks to GB for polishing the graphics.
We thank Rene's co-authors Andreas M. Antonopoulos and Olaoluwa Osuntokun for accepting Rene's decision to temporarily prioritize this research over their Lightning Network book project.
Finally our gratitude goes to Christian Decker with whom we had several discussions and who provided valuable feedback along the course of this research.
If you like the idea of decentralized and independent research and development of the Lightning Network feel free to support future work by contributing via \url{https://donate.ln.rene-pickhardt.de}.

\bibliography{mppSplitting}
\bibliographystyle{plain}

\begin{appendix}
  \section{Twitter-based TL;DR}
  While putting all together we actually shared the method and results in 6 Tweets:
  \\\\

    \textbf{March 17th 2021}\footnote{\url{https://twitter.com/renepickhardt/status/1372169686251626499}} \\
    Over the last year I have been making quite some discoveries about  \#bitcoin Payment Pathfinding on the \#LightningNetwork .\\A paper which introduces a probabilistic approach of modeling the uncertainty of remote channel balances is out and discussed on \url{https://lists.linuxfoundation.org/pipermail/lightning-dev/2021-March/002984.html}\\\\
  
  \textbf{April 22nd 2021}\footnote{\url{https://twitter.com/renepickhardt/status/1385144337907044352}} \\
  Multipathfinding 
\begin{enumerate}
\item capacities are integers (Satoshis)
\item channel success probabilities -log(1-x/c) are convex functions
\item Solving integer minimum cost flows with separable convex cost objective polynomially: \url{https://link.springer.com/chapter/10.1007\%2FBFb0121104}
\end{enumerate}
  Kudos @stefanwouldgo 4 digging this out
\\\\
  
  \textbf{May 26th 2021}\footnote{\url{https://twitter.com/renepickhardt/status/1397559345139888137}}\\
  most likely MPP-split to deliver 92 mBTC from my lightning node (03efc...) to (022c6...).
  We split into 11 onions!
\begin{figure}[htpb]
  \center
  \includegraphics[width=0.45\textwidth]{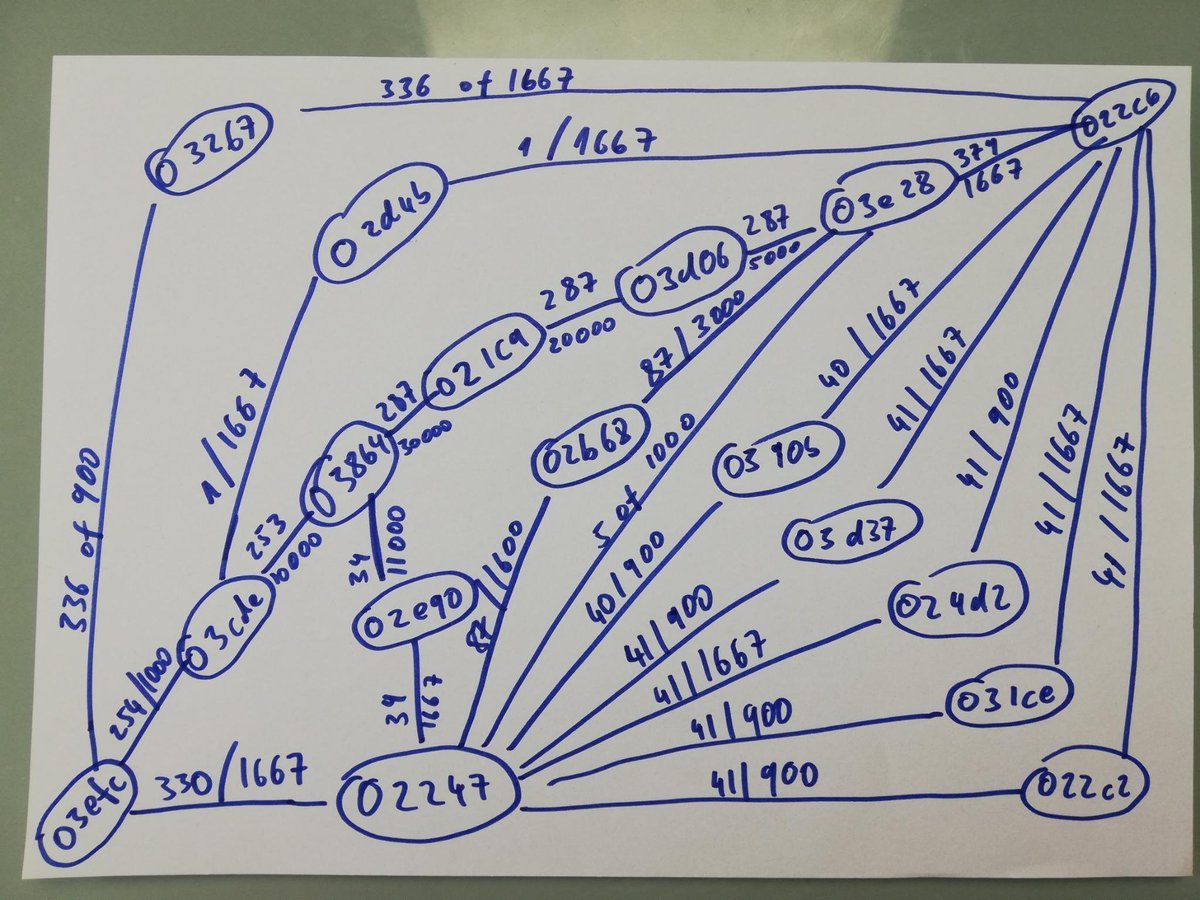}
  \label{fig:optflow}
\end{figure}\\
With knowledge of my own balance \& routing hints in the invoice the total likelihood to deliver all 11 onions is $64.84\%$
Note the 6 hop path with 4 WUMBO channels!
\\\\

\textbf{June 6th 2021}\footnote{\url{https://twitter.com/renepickhardt/status/1401514950984712198}}\\
Couldn't go to Miami so I coded up my algorithm!\\\\
In a simulated network with realistic balance values a node having 0.42 BTC could send 0.4 BTC to a remote node (that could receive up to 1.59 BTC) with no direct channel\\\\
It took just 4 attempts to deliver and 814 sat in fees\\\\

\textbf{June 8th 2021}\footnote{\url{https://twitter.com/renepickhardt/status/1402264479677693958}}\\
on other good news:
\begin{enumerate}
\item this yields another test vector for unit tests
\item the previous / non optimal algorithm might for other reasons (that go beyond a tweet) actually be better suitable for the lightning network after all
\item either way the issue seems fixable (:
\end{enumerate}

  \textbf{July 5th 2021}\footnote{\url{https://twitter.com/stefanwouldgo/status/1412158904008646660}}\\
  Lightning Routing IS NP hard, though.

  It's funny 'cause it's true.

\end{appendix}

%\end{multicols}
\end {document}